\documentclass[aps,nofootinbib,prd,eqsecnum,showkeys,reprint]{revtex4-2}
\usepackage{amsmath,graphicx}
\usepackage[colorlinks,linkcolor=blue]{hyperref}

\usepackage{physics,bm,booktabs,caption}

\begin{document}
	
\captionsetup[figure]{labelfont={bf},labelformat={default},labelsep=space,name={Fig.}}
\captionsetup[table]{labelfont={bf},labelformat={default},labelsep=space,name={Table}}
	
\title{Modified hybrid inflation in no-scale SUGRA with suppressed $R$-symmetry breaking}

\author{Qian Wan}
\email{wanqian@pku.edu.cn}
\author{Da-Xin Zhang}
\email{dxzhang@pku.edu.cn}
\date{\today}
\affiliation{School of Physics, Peking University, Beijing 100871, China}

\begin{abstract}
	A well-motivated cosmological hybrid inflation scenario based on no-scale SUGRA is considered.  It is demonstrated that an extra suppressed $R$-symmetry breaking term $S^n$ with $n\geq 4$ needs to be included in order to realize successful inflation. 
	The resulting potential is found to be similar (but not identical) to the one in the Starobinsky inflation model. A relatively larger tensor-to-scalar ratio $r\sim 10^{-2}$ and a spectral index $n_s\approx 0.965$ are obtained, which are approximately independent of $n$.
\end{abstract}



\maketitle
\tableofcontents

\section{Introduction}
Over the past few years, experiments and observations have greatly enhanced our comprehension of elementary particle physics and astrophysics, confirming the success of the Standard Model of particle physics and cosmology with remarkable precisions.
Nevertheless, there are still many challenges remain to be addressed, including the well-known flatness and horizon problems, the absence of cosmological relics, and the origin of the large scale structure.
As a well motivated framework to address these remaining challenges, the inflation scenario \cite{guthInflationaryUniversePossible1981,lindeNEWINFLATIONARYUNIVERSE1982,lythParticlePhysicsModels1999} is strongly supported in light of the observations of the cosmic microwave background radiation (CMBR) data from Planck \cite{planckcollaborationPlanck2018Results2020} and WMAP \cite{bennettNINEYEARWILKINSONMICROWAVE2013}.

Among the inflationary scenarios, the hybrid inflation models \cite{lindeHybridInflation1994,copelandFalseVacuumInflation1994,lindeHybridInflationSupergravity1997,rehmanHybridInflationRevisited2009} are especially attractive, since they can be naturally integrated into the grand unified theories (GUTs) through the introduction of a GUT gauge symmetry, where the inflaton is coupled to the GUT Higgs fields.
In addition, supersymmetry (SUSY) has been extensively employed in the construction of inflationary models \cite{ellisCosmologicalInflationCries1982,ellisPrimordialSupersymmetricInflation1983,ellisFluctuationsSupersymmetricInflationary1983}. SUSY provides a natural mechanism for maintaining the hierarchy between the inflation scale and the Planck scale without fine-tuning.
On the other hand, SUSY is also very important in the particle physics models for
providing a natural candidate for the dark matter, realizing gauge coupling unification
and stabilizing the  electro-weak scale.

In SUSY models, $R$-symmetry plays a crucial role. For instance, $R$-symmetry is related to SUSY breaking mechanism according to the Nelson-Seiberg theorem \cite{nelsonSymmetryBreakingSupersymmetry1994}, and is also of significant importance in gravity-mediated SUSY breaking scenarios \cite{pallisGravitymediatedSUSYBreaking2019,pallisSUSYbreakingScenariosMildly2021}. Additionally, $R$-symmetry is the natural choice for implementing the ``false'' vacuum inflationary scenario \cite{dvaliLargeScaleStructure1994}.
However, an exact $R$-symmetry forbids gauginos from getting masses and thus  must be broken spontaneously or explicitly.
Thus the inflation models with $R$-symmetry breaking have attracted a lot of attentions  in the literature. For example, in Ref. \cite{civilettiSymmetryBreakingSupersymmetric2013,khalilInspiredInflationModel2019} $R$-symmetry is broken softly by adding  to the superpotential a Planck suppressed dimension four operator, while $R$-symmetry is exact at the renormalizable level.

Inflation models within the framework of global SUSY has been extensively studied in the literature \cite{dvaliLargeScaleStructure1994}. However, it is necessary to extend them to models with local symmetry, namely supergravity (SUGRA) theories.
From a theoretical standpoint, the dynamics of inflation occur at the early universe and are sensitive to ultraviolet (UV) scales, hence it is essential to include gravity and upgrade global SUSY to SUGRA.
On the other hand, we expect to generate observable gravity waves from inflation, which requires the inflaton $S$ traverse field range compatible with or larger than the Planck scale according to the Lyth bound \cite{lythBoundInflationaryEnergy1984,lythWhatWouldWe1997}
\begin{equation}
	\label{eq1}
	r\leq 0.08\left(\frac{\Delta S}{M_P}\right)^2,
\end{equation}
where $M_P\approx 2.4\times 10^{18}\,\mathrm{GeV}$ is the reduced Planck scale, and will be set to unity for the rest of our discussion unless otherwise stated.

However, realizing inflation in SUGRA models is a challenging task. This is primarily due to the scalar potential of SUGRA involves an exponential factor of fields, which generally results in the inflaton acquiring a mass comparable to the Hubble scale. Consequently the slow roll condition $\eta\ll 1$ is violated
in the absence of fine-tuning and symmetrical reasons. This is also known as the $\eta$-problem \cite{lythParticlePhysicsModels1999,copelandFalseVacuumInflation1994,stewartInflationSupergravitySuperstrings1995}.
 
Given that the $\eta$-problem originates from the exponential factor in the F-term potential, a straightforward approach would be to consider the D-term inflation (DI), and similar work has been realized in \cite{binetruyDtermInflation1996,halyoHybridInflationFkom1996}. However, in this paper, we concentrate on F-term hybrid inflation (FHI), necessitating the development of an alternative solution. 
Furthermore, it has been proposed in the literature \cite{yamaguchiSupergravitybasedInflationModels2011} that the $\eta$-problem can be circumvented by imposing certain conditions (or symmetries) on the K\"{a}hler potential and the superpotential. For instance, the use of a non-compact Heisenberg symmetry \cite{binetruyNoncompactSymmetriesScalar1987,antuschChaoticInflationSupergravity2009}, a Nambu-Goldstone shift symmetry \cite{kawasakiNaturalChaoticInflation2000,yamaguchiNewInflationSupergravity2001,braxShiftSymmetryInflation2005}, or no-scale SUGRA \cite{khalilInspiredInflationModel2019,ellisNoScaleSupergravityRealization2013,romaoStarobinskylikeInflationNoscale2017} has been put forth as a potential solution.
Among the aforementioned  solutions, we are particularly interested in the no-scale SUGRA, as it allows for the realization of a Starobinsky-like inflation, with the predicted inflation observables falling within the core of the allowed regions of Planck data.

The purpose of this paper is to construct a class of hybrid inflation models within the no-scale SUGRA framework.
The most general superpotential at the renormalizable level is $R$-symmetric, but we find an explicit $R$-symmetry breaking term beyond the renormalizable level is indispensable, which plays an important role in realizing successful inflation.
The organization of this paper is as follows.
In Section \ref{section2} we present  the simplest SUSY hybrid model with $R$-symmetry, and study how it can be modified to realize a Starobinsky-like inflation in the no-scale SUGRA.
Section \ref{section3} provides a detailed analysis of the inflation predictions.
Section \ref{section4} is devoted to the conclusion, some brief discussions about monopole problems are also included.

\section{The model}
\label{section2}
\subsection{General discussion}
We start by considering the simplest SUSY hybrid inflation model presented in Ref. \cite{dvaliLargeScaleStructure1994,pallisInducedgravityGUTscaleHiggs2018,pallisStarobinskyTypeBLHiggs2023}. In this paper the inflaton $S$ is a gauge singlet with the $R$ charge 2, and the trigger fields $\phi$, $\bar{\phi}$ with $R$ charge 0 denote the Higgs superfields transforming as non-trivial conjugate representations of the GUT gauge group $G$.
The most general renormalizable superpotential that preserves gauge symmetry and $R$-symmetry is given by
\begin{equation}
	\label{eq2}
	W=\kappa S(\bar{\phi}\phi-M^2),
\end{equation}
where $\kappa$ is a dimensionless parameter and can be taken to positive without loss of generality, since the phase of $\kappa$ can be absorbed into the redefinition of $S$-field, and
$M$ is the mass scale at which the GUT symmetry $G$ is broken to its subgroup at the end of inflation, and we will take $M\approx 2.4\times 10^{16}\,\mathrm{GeV}$ in the subsequent discussion.

However, similar to \cite{khalilInspiredInflationModel2019}, this simple superpotential \eqref{eq2} does not result in a slowly rolling scalar potential within the framework of no-scale SUGRA. Therefore an additional term $S^n(\bar{\phi}\phi)^m$ respecting the gauge symmetry must be included in the superpotential, and it is evident that such a term with $n\neq 1$ must violate the $R$ symmetry.
Given that we are focusing on standard inflationary trajectory $|\phi|=|\bar{\phi}|=0$, additional terms with $m\neq 0$ will not affect the dynamics of inflation, thus we will select $m=0$ for the sake of simplicity.
It should be noted that these terms with $m\neq 0$ will play an important role in the shifted hybrid inflation models, although this is not discussed in this paper.
With regard to the parameter $n$, it will be demonstrated subsequently that a Starobinsky-like scalar potential is available only when $n\geq 4$, and the natural requirement $\kappa\sim\order{0.1}$ leads us to choose $n=4$.
It is worth noting that a Starobinsky-like potential can also be realized by including both $S^2$ and $S^3$ terms in Ref. \cite{romaoStarobinskylikeInflationNoscale2017,moursyNoscaleHybridInflation2021}.

Including an explicit $R$-symmetry breaking term $S^n$ with $n\geq 4$, the general superpotential is
\begin{equation}
	\label{eq3}
	W=\kappa S(\bar{\phi}\phi-M^2)+\frac{\lambda}{n}\frac{S^n}{M_*^{n-3}},
\end{equation}
where  the dimensionless coupling $|\lambda|\ll 1$ is responsible for the $R$-symmetry breaking
and can be in general complex, but we assume it to be real and positive similar to $\kappa$ for the sake of simplicity.
$M_*$ is a mass scale whose natural choice  would be the reduced Planck scale $M_P$, given that a global symmetry is expected to be broken by the gravitational effects.
It is noteworthy that the $R$-symmetry is recovered in the limit $\lambda\to 0$, which indicates that our model is still natural in 't Hooft's  sense \cite{Hooft1980RecentDI}.

We consider the following K\"{a}hler potential with no-scale symmetry in order to avoid the $\eta$-problem and realize a Starobinsky-like inflation,
\begin{equation}
	\label{eq4}
	K=-3\ln\left[T+T^*-\frac{|S|^2}{3}-\frac{|\phi|^2}{3}-\frac{|\bar{\phi}|^2}{3}\right],
\end{equation}
where $T$ is the modulus field. The Lagrangian of the complex scalar fields, determined by the K\"{a}hler potential $K$ and the superpotential $W$, can be expressed as
\begin{equation}
	\mathcal{L}=K_i^j\partial_\mu\phi^i\partial^\mu\phi_j^*-V(\phi^i,\phi_i^*),
\end{equation}
where $K_i^j\equiv\partial^2K/\partial\phi^i\partial\phi_j^*$ is the K\"{a}hler metric. The potential $V$ of scalar fields $\phi^i$ is consist of two parts, the $F$-term potential $V_F$ and the $D$-term potential $V_D$.
The $F$-term potential $V_F$ is given by
\begin{equation}
	V_F=\mathrm{e}^K\left[D_iW(K^{-1})_j^iD^jW^*-3|W|^2\right]
\end{equation}
with
\begin{equation}
	D_iW=\frac{\partial W}{\partial\phi^i}+W\frac{\partial K}{\partial\phi^i},\quad D^jW^*=\frac{\partial W^*}{\partial\phi_j^*}+W^*\frac{\partial K}{\partial\phi_j^*},
\end{equation}
where $\phi^i\in\{T,S,\phi,\bar{\phi}\}$ represents the scalar component of the corresponding superfield. The $D$-term potential $V_D$ is related to the gauge symmetry and is given by
\begin{equation}
	V_D=\frac{g^2}{2}\Re f_{ab}^{-1}D^aD^b
\end{equation}
with
\begin{equation}
	D^a=\phi^i(T^a)_i^j\frac{\partial K}{\partial\phi^j}+\xi^a,
\end{equation}
where $f_{ab}(\phi^i)$ is the gauge kinetic function, $g$ is the gauge coupling constant determining the strength of the gauge interactions, and $T^a$ are generators of the GUT gauge group in the associated representation. The so-called Fayet-Iliopoulos (FI) terms $\xi^a$ contributes to the $D$-term potential only when the gauge symmetry is $U(1)$.

It is acknowledged that FI terms play a significant role in D-term inflation \cite{binetruyFayetIliopoulosTermsSupergravity2004}. However, in the present study, FI terms have been set to zero by default. This choice is based on three main reasons. Firstly, the gauge group $G$ does not in general contain the $U(1)$ factor, in which case the FI terms do not contribute to the D-term. Secondly, for the case where $G$ contains $U(1)$, we can ignore the contributions of the FI terms as well, since within a reasonable parameter range the contributions of the FI terms can always be considered as a negligible perturbations of the inflation potential without affecting our subsequent discussion on the cosmological observables \cite{heurtierSingleFieldInfation2015}. Thirdly, we have primarily focused on F-term hybrid inflation scenario in our discussion. Consequently, the D-flatness conditions are always satisfied, and the specific value of the FI terms do not affect our subsequent discussions.

Working in the $D$-flat direction $|\phi|=|\bar{\phi}|$, the total scalar potential will be given by the SUGRA $F$-term scalar potential (we use the same notations for superfields and their scalar components for simplicity),
\begin{equation}
	\label{eq5}
	 V_F=\frac{1}{\Omega^2}\sum_i\left|W_i\right|^2,\quad\Omega
=T+T^*-\frac{|S|^2}{3}-\frac{|\phi|^2}{3}-\frac{|\bar{\phi}|^2}{3}.
\end{equation}
Following Ref. \cite{ellisNoScaleSupergravityRealization2013}, it is assumed that the modulus field $T$ is stabilized at a fixed scale such that\footnote{The value of $\expval{T}$ is inconsequential and the discussions of its stabilization can be found in  \cite{kachruSitterVacuaString2003,balasubramanianSystematicsModuliStabilisation2005}.} $\expval{T}=\expval{T^*}=1/2$. This stabilization requires a nonperturbative effect at a high scale. It is evident that the scalar potential described above is positive semi-definite. Consequently, it possesses two global minimum which are supersymmetric and Minkowskian, located at
\begin{equation}
	S=0,\quad|\phi|=|\bar{\phi}|=M.
\end{equation}
In addition to the above SUSY vacua, there are also other global minima located at $\phi=0$ and $S=(\kappa M^2/\lambda)^{1/n-1}$. However the inflaton will not slow roll to these vacua as we will be demonstrated later.
It is worth noting that $S=0$ and $|\phi|=|\bar{\phi}|=0$ is a false vacuum, so we cannot produce a successful inflationary scenario without the trigger fields. This is quite different from  \cite{khalilInspiredInflationModel2019} where the inflaton $S$ rolls to the origin.

It is obvious that the potential \eqref{eq5} is minimized along the $D$-flat direction $|\phi|=|\bar{\phi}|=0$, so the Higgs fields are fixed at the origin during the inflation, resulting in an effective inflationary potential given by
\begin{equation}
	V_{F}=\frac{1}{(1-|S|^2/3)^2}|-\kappa M^2+\lambda S^{n-1}|^2,
\end{equation}
which is simlilar to the potential obtained in  \cite{khalilInspiredInflationModel2019,ellisNoScaleSupergravityRealization2013,romaoStarobinskylikeInflationNoscale2017}.
To better study the potential for the inflation, we redefine the $S$ field in terms of $\chi$ following \cite{ellisNoScaleSupergravityRealization2013},
\begin{equation}
	\label{eq6}
	S=\sqrt{3}\tanh\left(\frac{\chi}{\sqrt{3}}\right),\quad\chi=\frac{x+\mathrm{i} y}{\sqrt{2}},
\end{equation}
where $x$ is the canonically normalized inflaton. Then the Lagrangian becomes
\begin{equation}
	\begin{aligned}
		\mathcal{L}=&\sec^2\left(\sqrt{\frac{2}{3}}y\right)\bigg[\frac{1}{2}(\partial_\mu x)^2+\frac{1}{2}(\partial_\mu y)^2 \\
		&-\kappa^2M^4\left|\cosh^2\frac{x+\mathrm{i} y}{\sqrt{6}}\left(1-a\tanh^{n-1}\frac{x+\mathrm{i} y}{\sqrt{6}}\right)\right|^2\bigg],
	\end{aligned}
\end{equation}
where $a=\sqrt{3^{n-1}}\lambda/\kappa M^2$ is a dimensionless parameter.
In the specific case where $a=1$ (the reasons will become apparent late), the mass squared of the $y$-field is $m_y^2=\kappa^2M^4/3$ during inflation when $x$ is large and $m_y^2=\kappa^2M^4(n-1)^2/3$ at the end of inflation when $x=0$. It is evident that the mass squared $m_y^2$ is larger than the Hubble scale during inflation (a numerical estimate will be provided later), therefore  $y$ will be frozen at the origin during inflation.
It is therefore safe to set $y=0$ and then the kinetic terms for $x$ and $y$ become canonical. Finally, an effective single field inflation is obtained. The $F$-term scalar potential responsible for inflation will be given by
\begin{equation}
		\label{eq7}
		 V_{\text{inf}}^{(n)}=\kappa^2M^4\cosh^4\left(\frac{x}{\sqrt{6}}\right)\left(1-a\tanh^{n-1}\left(\frac{x}{\sqrt{6}}\right)\right)^2.
\end{equation}

Now we turn to discuss the conditions necessary to obtain a Starobinsky-like potential $V\sim (1-\mathrm{e}^{-x})^2$. To this end, we will consider the limit $x\to\infty$,  where the effective potential will take the following form
\begin{equation}
	\label{eq7a}
	\begin{aligned}
		V_{\text{inf}}^{(n)}\approx&\frac{\kappa^2M^4}{16}\bigg[(1-a)\mathrm{e}^{\sqrt{2/3}x}+2(1-2a+na) \\
		&+(1-7a+8na-2n^2a)\mathrm{e}^{-\sqrt{2/3}x}\bigg]^2.
	\end{aligned}
\end{equation}
\begin{figure}[ht]
	\centering
	\includegraphics[scale=0.5]{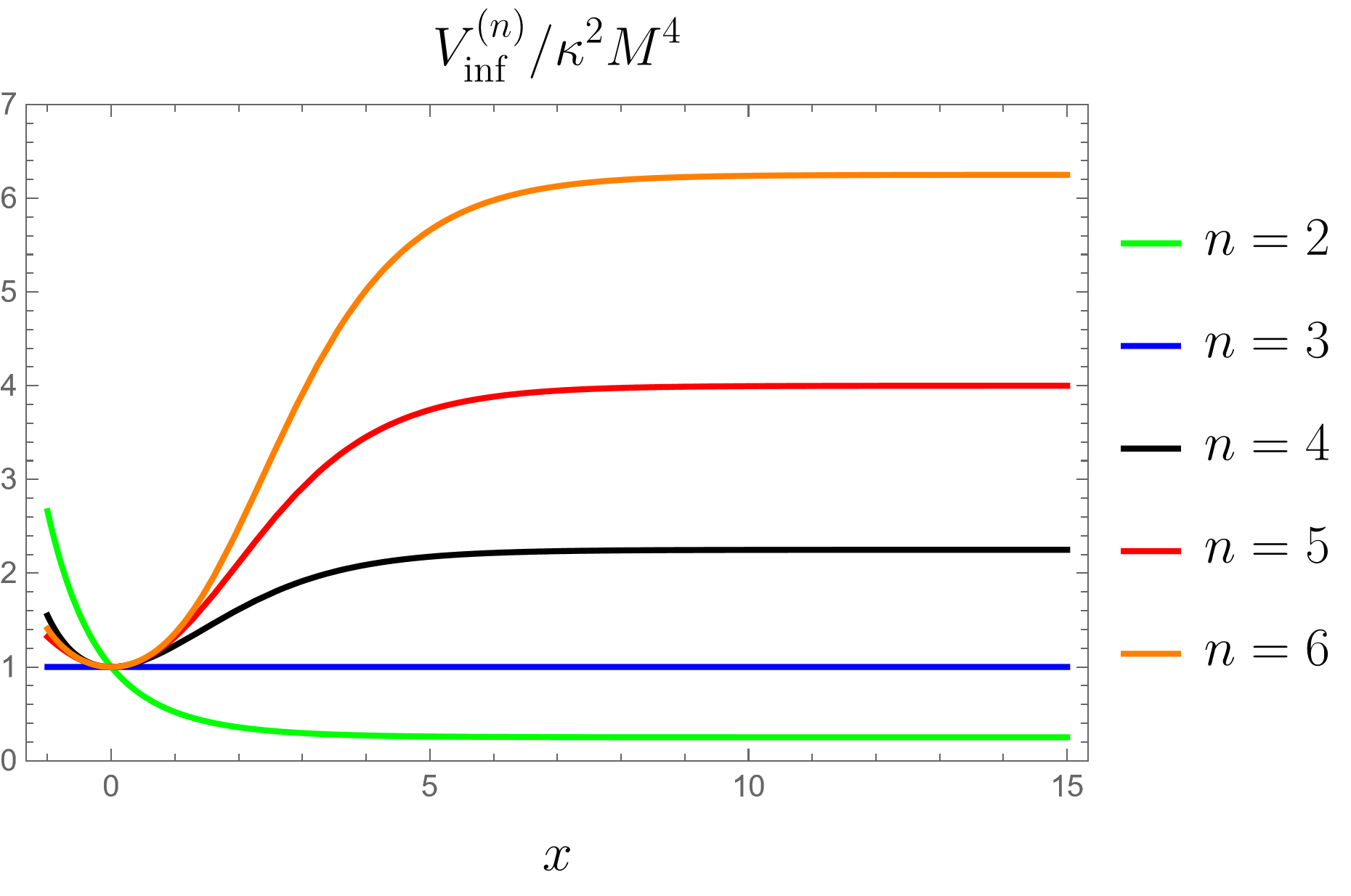}
	\caption{The effective inflationary potential \eqref{eq7} $V_{\text{inf}}^{(n)}/\kappa^2M^4$ for different values of $n$ with $a=1$}
	\label{Fig1}
\end{figure}
It is clear that the parameter $a$ must be fine-tuned to unity in order to ensure the potential stays flat in the limit $x\to\infty$, which is the characteristic of Starobinsky-like potential.
By taking $a=1$, the potential (normalized by $\kappa^2M^4$) for different values of $n$ is depicted in Fig. \ref{Fig1}.
It is demonstrated that the corresponding scalar potential of $n=2$ (denoted by the green curve in Fig. \ref{Fig1}) and $n=3$ (denoted by the blue curve in Fig. \ref{Fig1}) are not Starobinsky-like and do not result in successful slow roll inflation.
Furthermore, it can be seen from \eqref{eq7a} that the dimensionless potential $V_{\text{inf}}^{(n)}/\kappa^2M^4\approx (n-1)^2/4$ increases as $n$ becomes larger, indicating that a smaller value of $\kappa$ is necessary to keep $[V_{\text{inf}}^{(n)}]^{1/4}$ fixed at the inflationary scale. It can be demonstrated later that the maximum value of the parameter $\kappa\sim\order{0.1}$ is obtained for $n = 4$, which also represents the lowest operator at the non-renormalizable level.

\subsection{The case with $n=4$}
In order to obtain a large $\kappa$ close to unity that satisfies naturalness, we consider the case of $n=4$, where the superpotential remains $R$-symmetrical at the renormalizable level.
A similar superpotential has also been considered in   \cite{civilettiSymmetryBreakingSupersymmetric2013}, where a successful inflationary scenario is realized by employing the minimal K\"{a}hler potential and including the soft SUSY breaking term.
However, the tensor-to-scalar ratio $r\sim 10^{-8}$ is so small that it is unlikely to be observed by current or future experimental endeavors, though this can be cured by adding higher-order terms to the K\"{a}hler potential \cite{shafiObservableGravityWaves2011}.
It is also worth noting that allowing the non-renormalizable $R$-symmetry breaking terms in the superpotential is a simple solution to generate right-handed neutrino masses in flipped $SU(5)$ model, as discussed in  \cite{civilettiSymmetryBreakingSupersymmetric2013}.

\begin{figure}[ht]
	\centering
	\includegraphics[scale=0.5]{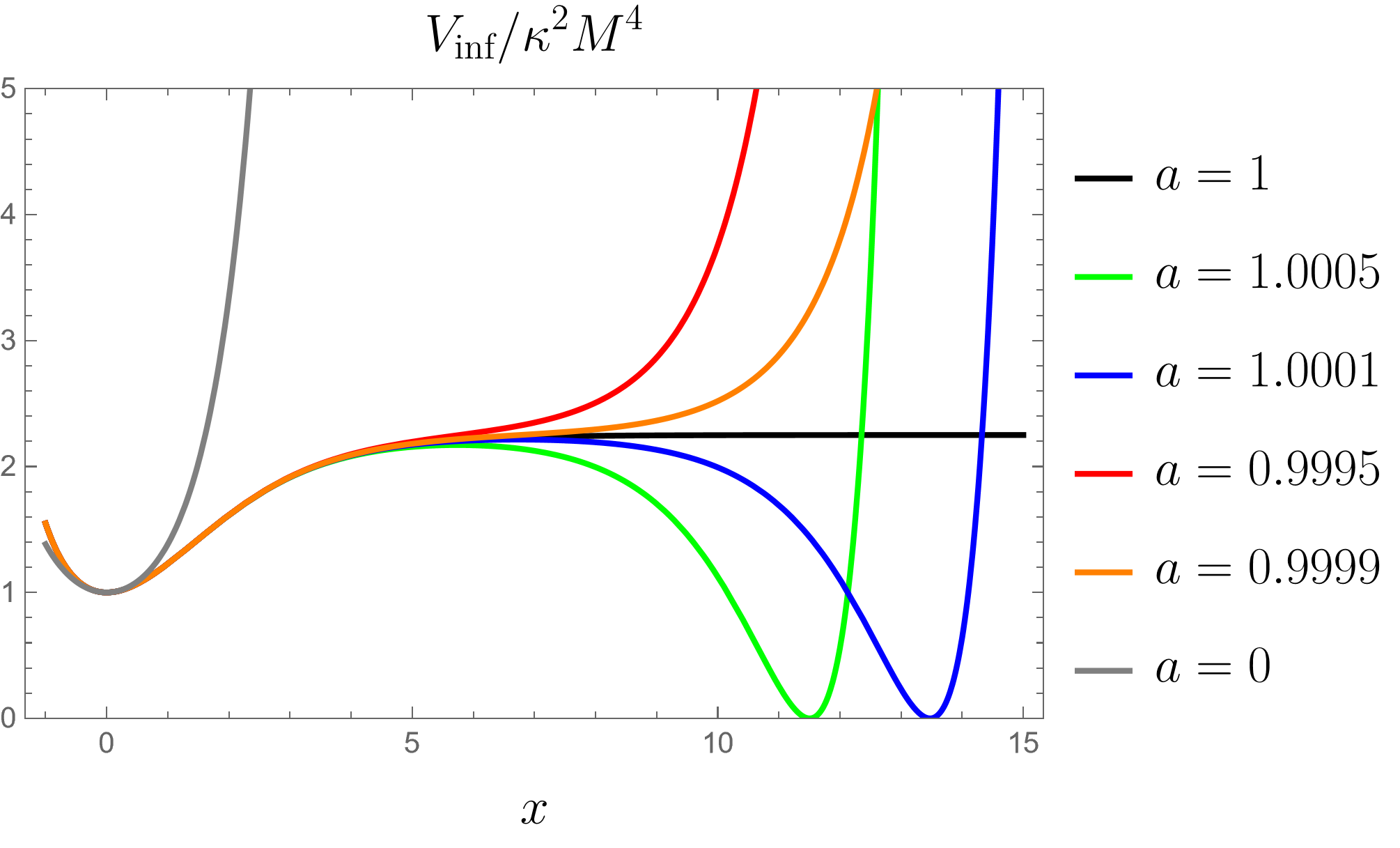}
	\caption{The effective inflationary potential \eqref{eq5a} $V_{\text{inf}}/\kappa^2M^4$ with $n=4$ for different values of $a$}
	\label{Fig4}
\end{figure}

Taking $n=4$ in \eqref{eq7}, the effective scalar potential is given by
\begin{equation}
	\label{eq5a}
	V_{\text{inf}}^{(4)}=\kappa^2M^4\cosh^4\left(\frac{x}{\sqrt{6}}\right)\left(1-a\tanh^3
	\left(\frac{x}{\sqrt{6}}\right)\right)^2,
\end{equation}
and the potential (normalized by $\kappa^2M^4$) for different values of $a$ is shown in Fig. \ref{Fig4}. In addition, the potential for $a=0$ is displayed in order to illustrate the importance of the $R$-symmetry breaking term.

In the limit $a=1$, the effective inflationary potential $V_{\text{inf}}$ (denoted by the black line in Fig. \ref{Fig4}) is Starobinsky-like and becomes flat for large values of $x$. A small variation of $a$ from unity can lead the potential become steep, and it is expected that for any significant deviation of $a$ from unity, the slow roll of the inflaton is spoiled.
In the case  $a=0$ (i.e., $\lambda=0$), the effective potential (denoted by the gray curve in Fig. \ref{Fig4}) is too steep, indicating that such a potential cannot provide sufficient inflation.
Hence the explicit $R$-symmetry breaking term plays an important role in the realization of successful inflation.

\begin{figure*}[ht]
	\centering
	\includegraphics[scale=0.65]{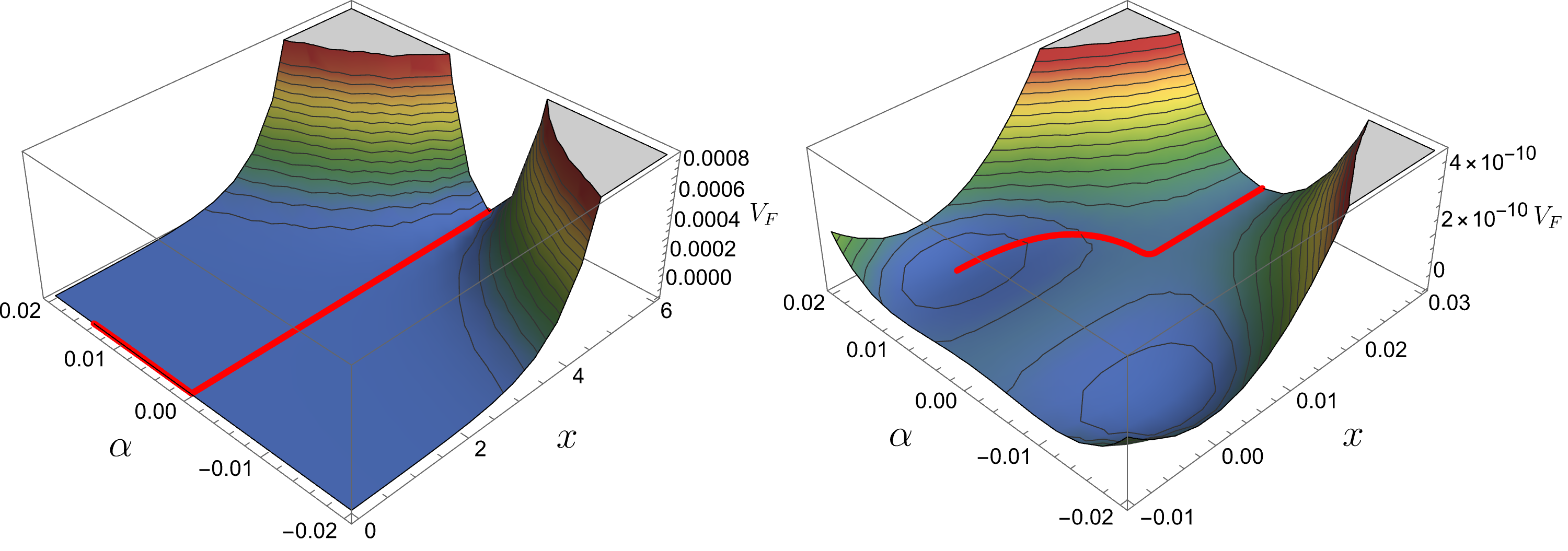}
	\caption{The $F$-term scalar potential of $x$ and $\alpha$ for large (left panel) and small (right panel) values of $x$, with $n=4$, $M=0.01$, $a=1$ and $\kappa=0.083$. All values are given in units where $M_P=1$. The inflationary trajectory has been plotted by red curve}
	\label{Fig2}
\end{figure*}

The total scalar potential \eqref{eq5} contains three complex scalar fields, one real degree of freedom cancels since $|\phi|=|\bar{\phi}|$ along the inflationary trajectory.
The complex scalar fields $\phi$, $\bar{\phi}$ can be parameterized in terms of their three real components following \cite{moursyNoscaleHybridInflation2021}
\begin{equation}
	 \phi=\frac{\alpha+\mathrm{i}\beta}{\sqrt{2}}\mathrm{e}^{\frac{\mathrm{i}\Sigma}{\sqrt{2}M}},\quad\bar{\phi}
=\frac{\alpha+\mathrm{i}\beta}{\sqrt{2}}\mathrm{e}^{\frac{-\mathrm{i}\Sigma}{\sqrt{2}M}},
\end{equation}
where $\Sigma$  corresponds to the massless Goldstone boson which  will be eaten by the gauge boson. Consequently, this degree of freedom will not contribute to the dynamics of inflation. We  plot the scalar potential \eqref{eq5} in Fig. \ref{Fig2}. Inflation occurs along the local minimum $|\phi|=|\bar{\phi}|=0$ of the potential as the inflaton $x$ rolls slowly toward the critical point $x_c$, then the field falls naturally into one of two global minima at $|\phi|=M$ ($|\alpha|=\sqrt{2}M$).

\begin{figure*}[ht]
	\centering
	\includegraphics[scale=0.5]{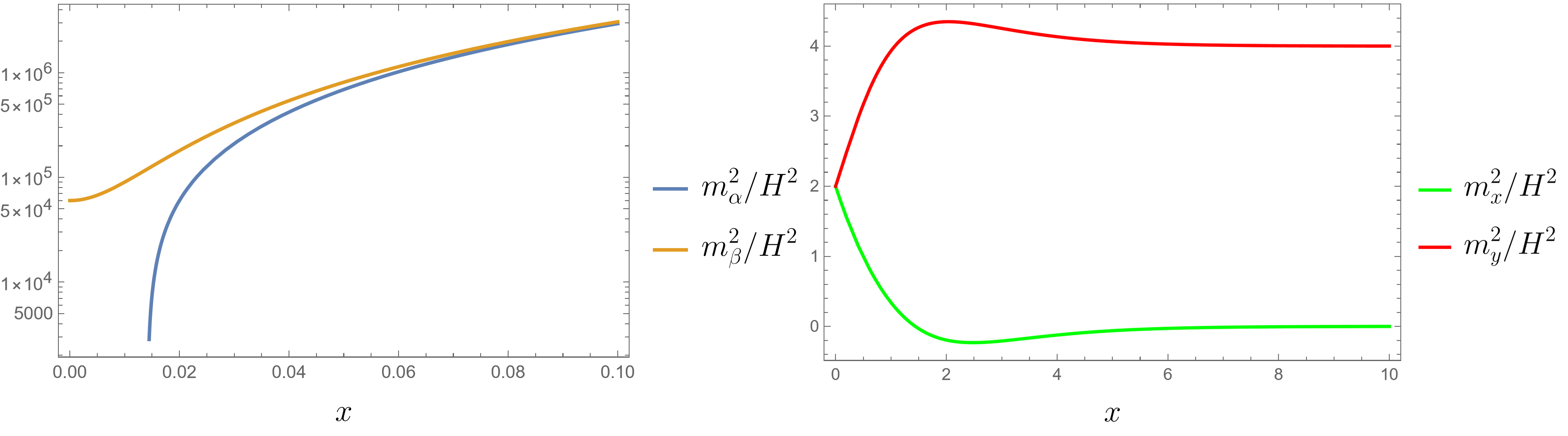}
	\caption{The ratios of the squared mass of the real scalar fields $\alpha$, $\beta$, $y$ and $x$, to the squared Hubble scale, with $n=4$, $M=0.01$, $a=1$ and $\kappa=0.083$}
	\label{Fig3}
\end{figure*}

Now we return to discuss the stabilization of the non-inflaton fields. The mass-squared matrices $M_{ij}=\partial^2V/\partial\phi_i\partial\phi_j$ of all the real scalar fields along the inflationary trajectory have been calculated, and the ratios of field dependent squared masses to the squared Hubble scale during inflation are plotted in Fig. \ref{Fig3}.
It is clear that $\alpha$, $\beta$ and $y$ acquire masses larger than the Hubble scale given by $H^2=V_\text{inf}/3$ during inflation, hence they are frozen at the origin during the inflation without affecting the inflation dynamics.
This finally results in a simple effective single-field inflation, which circumvents the occurrence of unacceptably large isocurvature fluctuations in the multi-field inflation scenarios.

Fig. \ref{Fig3} also illustrates that as the inflaton $x$ rolls down, the value of $m_\alpha^2$ decreases
and changes to negative at a critical point $x_c$.
In particular, for small $x$, the leading order approximation of $m_\alpha^2$ is given by
\begin{equation}
	m_\alpha^2\approx-\frac{2\kappa^2M^2}{3}(3-2M^2)+\frac{\kappa^2}{3}\left(3-2M^2+2M^4\right)x^2,
\end{equation}
where $M\ll 1$ is the GUT scale, therefore we always have $m_\alpha^2<0$ for small values of $x$. The critical point $x_c$ is determined by taking $m_\alpha^2=0$ and we have
\begin{equation}
	x_c\approx\sqrt{\frac{6M^2-4M^4}{3-2M^2+2M^4}}\approx 0.014.
\end{equation}
When $x$ rolls to $x_c$, $\alpha=0$ becomes a local maximum and $\alpha$ goes to its true minimum $\alpha=\sqrt{2}M$ as indicated in Fig. \ref{Fig2}, where the inflationary trajectory (red curve) has been plotted. On the other hand, $y$ and $\beta$ are fixed at zeros during and after inflation.

Considering that SUSY is broken along the inflationary trajectory, we now turn to discuss the contributions from radiative corrections to the tree-level scalar potential for consistency.
In the one loop approximation, the effective potential is given by \cite{colemanRadiativeCorrectionsOrigin1973}
\begin{equation}
	V_\text{1-loop}=\frac{1}{64\pi^2}\mathrm{Str}\,M_i^4\ln\frac{M_i^2}{\Lambda^2},
\end{equation}
where the supertrace is taken over all superfields with inflaton dependent masses and $\Lambda$ is a cutoff scale. As previously proposed, the stabilized fields during inflation have masses $M_i\sim H$. Given that $H^2=V_{\text{inf}}/3\sim 10^{-10}$, the one loop correction $V_{\text{1-loop}}\sim H^4/64\pi^2\sim 10^{-22}$ which is negligible compared to the tree level potential, and thus will be ignored in the rest of our discussion.

\section{Results and discussions}
\label{section3}
In this section we will proceed to determine the inflationary observables and find the constraints on different parameters.
To solve the fundamental cosmological equations, the slow-roll approximation will be used throughout, whereby inflation occurs while the slow-roll parameters are less than unity.
In the $M_P=1$ units, the slow-roll parameters are expressed as
\begin{equation}
	\epsilon=\frac{1}{2}\left(\frac{V'}{V}\right)^2\,,\quad\eta=\frac{V''}{V},
\end{equation}
where the primes denote the derivatives with respect to the canonical inflaton $x$. Inflation ends either when one of the slow-roll parameters become unity or when the inflaton field reaches the waterfall point $x_c$.

In the slow roll approximation (i.e. $\epsilon,\,|\eta|\ll 1$), the number of e-foldings $N_*$ is given by
\begin{equation}
N_*=\int_{x_e}^{x_*}\frac{1}{\sqrt{2\epsilon}}\mathrm{d} x,
\end{equation}
where $x_*$ represents the value of the inflaton field when the pivot scale $k_*$ exits the horizon, and $x_e$ denotes the field value at the end of inflation. The value of $N_*$ depends on the energy scale during inflation and cosmic history after inflation, and is typically taken to be 50 or 60.

The curvature perturbations are generated through the inflaton fluctuations during inflation. The scalar spectral index $n_s$, the tensor to scalar ratio $r$ and the amplitude of the curvature perturbation $A_s$ are given by
\begin{equation}
\label{eq8}
n_s\approx 1-6\epsilon+2\eta,
\end{equation}
\begin{equation}
r\approx 16\epsilon,
\end{equation}
and
\begin{equation}
A_s\approx \frac{V}{24\pi^2\epsilon},
\end{equation}
where $A_s$ is normalized to $2.137\times 10^{-9}$ at the pivot scale $k_*=0.05\,\mathrm{Mpc}^{-1}$ according to Planck 2018 data \cite{planckcollaborationPlanck2018Results2020}.

Table \ref{tab1} presents the predicted values of our modified model for varying values of $n$, with $a=1$ and $N_*=60$.
It can be observed that the parameters $\kappa$ and $\lambda$ exhibit  decreasing trends as $n$ increases, which further substantiates the previous conclusion.
Additionally, the initial value of the inflaton field $x_*$ demonstrates an upward trend as $n$ increases, and it is notable that the predicted values of the inflation observations $n_s\approx 0.961$ and $r\approx 0.005$ corresponding to different $n$ are approximately fixed.

\begin{table}[ht]
	\centering
	\caption{Inflation predictions for different values of $n$ with $N_*=60$ and $a=1$.}
	\label{tab1}
	\begin{tabular}{cccccc}
		\hline
		$n$ & $\kappa$ & $\lambda$ & $x_*$ & $n_s$ & $r$ \\
		\hline
		$4$ & $0.083$ & $1.596\times 10^{-6}$ & $5.119$ & $0.960650$ & $0.004747$ \\
		$5$ & $0.061$ & $6.792\times 10^{-7}$ & $6.002$ & $0.960911$ & $0.004587$ \\
		$6$ & $0.049$ & $3.126\times 10^{-7}$ & $6.507$ & $0.960915$ & $0.004555$ \\
		$7$ & $0.041$ & $1.504\times 10^{-7}$ & $6.861$ & $0.960838$ & $0.004558$ \\
		$8$ & $0.035$ & $7.448\times 10^{-8}$ & $7.135$ & $0.960782$ & $0.004562$ \\
		\hline
	\end{tabular}
\end{table}

\begin{figure}[ht]
	\centering
	\includegraphics[scale=0.6]{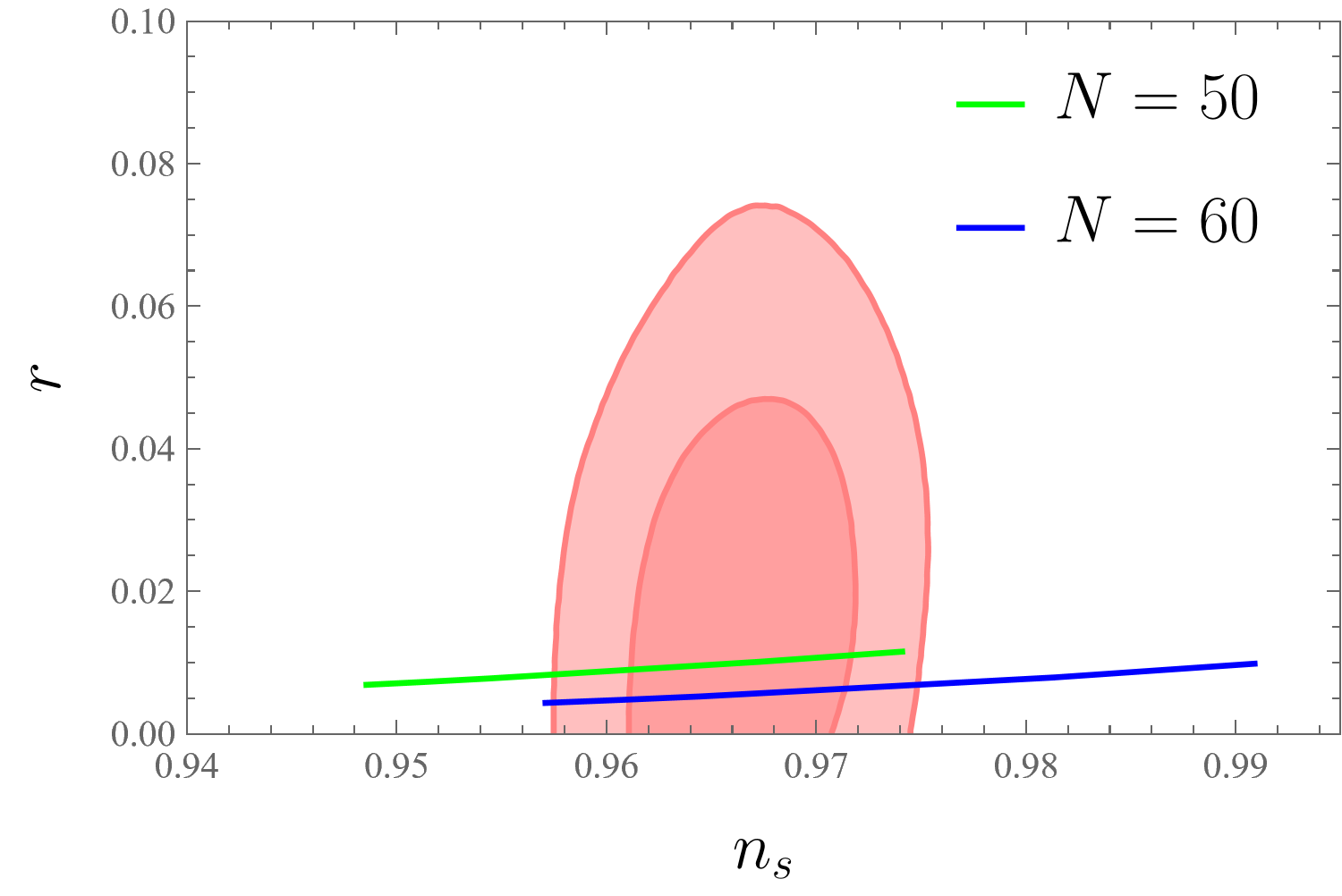}
	\caption{The predictions of our no-scale SUGRA model with $n=4$ for the spectral index $n_s$ and the tensor-to-scalar ratio $r$, compared with the 68\% and 95\% C.L. regions found in analyses of Planck 2018 dataset (Planck TT + low E + BKP + BAO +lensing) \cite{planckcollaborationPlanck2018Results2020}. Here we scanning over $a\in[1.0001,0.9993]$ with $N=50$ (green line) and $N=60$ (blue line)}
	\label{Fig5}
\end{figure}

For the case of $n=4$, we displayed the predictions of our modified model for $n_s$, $r$ in a plot versus the Planck limits in Fig. \ref{Fig5}.
In these numerical calculations, we have imposed the constraint that the number of e-foldings $N_*=50\sim 60$.

As illustrated in Fig. \ref{Fig5}, inflation predictions within the observed range of Planck data can be obtained by $a$ varying from 0.9993 to 1.0001.
By setting $a=1$ and $N_*=60$, we obtain $n_s\approx 0.961$ and $r\approx 0.0048$ which fall within the allowed range of the Planck data.
Meanwhile, the corresponding parameters in the superpotential are $\kappa\approx 0.083$ and $\lambda\approx 1.596\times 10^{-6}$. It is evident that a tiny value of $\lambda$ is necessary to facilitate a slow-roll inflationary process, which is also in line with the supposition that the violation of $R$-symmetry is significantly suppressed.
We also find that a large field value $x_*\sim 5$ (for typical valule of $a=1$ and $N_*=60$, we obtained $x_*=5.119$) in our model result in the inflaton amplitude $\Delta x\sim 5$ to be super-Planckian and the upper bound of \eqref{eq1} has been raised, this explain why the tensor-to-scalar ratio $r$ in our model is larger than  that in  \cite{civilettiSymmetryBreakingSupersymmetric2013} with a similar superpotential. 
It is also easy to observed from Fig. \ref{Fig4} that the initial value $x_*$ is less than the local maximum point of the potential, hence the inflaton will evolve to the critical point $x_c$ rather than the global minimum at $S=(\kappa M^2/\lambda)^{1/3}$.

\section{Conclusion}
\label{section4}
Before concluding we would like to give a brief comment on the monopole problem, since the topological defects are often produced in quantities when the gauge symmetry is broken, which is in contradiction with the fact that no magnetic monopoles have been observed. 
There are two potential ways to solve this problem.
One solution is to select a shifted inflationary trajectory $|\phi|=|\bar{\phi}|\neq 0$ such that the gauge symmetry is broken before the end of inflation. This can be achieved by introducing higher order terms $S^n(\bar{\phi}\phi)^m$ with $m\neq 0$, and similar works can be found in \cite{khalilInflationSupersymmetricSU2011} with the minimal K\"{a}hler potential or \cite{ahmedObservableGravitinoDark2023,ijazExploringPrimordialBlack2024} with the no-scale K\"{a}hler potential.
Given that our work is discussed in the standard track, we prefer to choose the second solution, namely the selection of a gauge group that does not give rise to topological defects in the context of gauge symmetry breaking. One particularly compelling gauge group is the flipped $SU(5)$, which  precludes the formation of stable monopoles. Once we choose the gauge symmetry group, the reheating process can be studied as well, see \cite{kyaeFlippedSUPredicts2006} for more information.

In conclusion, we have discussed in details how to realize hybrid inflation within the framework of no-scale SUGRA.
The simplest superpotential with $R$-symmetry does not result in successful inflation, since the associated scalar potential is extremely steep.
Extra $R$-symmetry breaking terms $S^n(\bar{\phi}\phi)^m$ with $n\geq 4$ are needed to be included in order to achieve the Starobinsky-like inflation, which is strongly favored by the recent observation data from the Planck satellite.
In comparison to the flipped $SU(5)$ model with the minimal K\"{a}hler potantial in \cite{civilettiSymmetryBreakingSupersymmetric2013}, we have obtained a relatively large values of $r\sim 10^{-2}$ around $n_s\approx 0.965$, which could potentially be measurable by the future experiments.
Such modified models with explicit $R$-symmetry breaking at the non-renormalizable level may also have some interesting consequences. For example, they provide a simple solution to generate the masses of right-handed neutrinos in the flipped $SU(5)$ model, and also provide a mass for $R$-axion in dynamical supersymmetry breaking model \cite{baggerRaxionDynamicalSupersymmetry1994}.

\bibliography{mylibrary}  

\end{document}